\documentclass[pra, twocolumn, showpacs, floatfix, superscriptaddress]{revtex4}

\usepackage{times}
\usepackage{amsfonts}
\usepackage{amssymb}
\usepackage{amsmath}
\usepackage{graphicx}

\begin{document}

\title{Parametric amplification of matter waves in dipolar spinor Bose-Einstein condensates}

\author{F. Deuretzbacher}
\email{fdeuretz@itp.uni-hannover.de}
\affiliation{Institut f\"ur Theoretische Physik, Leibniz Universit\"at Hannover, Appelstr. 2, 30167, Hannover, Germany}

\author{G. Gebreyesus}
\affiliation{Institut f\"ur Theoretische Physik, Leibniz Universit\"at Hannover, Appelstr. 2, 30167, Hannover, Germany}

\author{O. Topic}
\affiliation{Institut f\"ur Quantenoptik, Leibniz Universit\"at Hannover, Welfengarten 1, 30167 Hannover, Germany}

\author{M. Scherer}
\affiliation{Institut f\"ur Quantenoptik, Leibniz Universit\"at Hannover, Welfengarten 1, 30167 Hannover, Germany}

\author{B. L\"ucke}
\affiliation{Institut f\"ur Quantenoptik, Leibniz Universit\"at Hannover, Welfengarten 1, 30167 Hannover, Germany}

\author{W. Ertmer}
\affiliation{Institut f\"ur Quantenoptik, Leibniz Universit\"at Hannover, Welfengarten 1, 30167 Hannover, Germany}

\author{J. Arlt}
\affiliation{QUANTOP, Danish National Research Foundation Center for Quantum Optics, Department of Physics and Astronomy, Aarhus University, Ny Munkegade 120, DK-8000 Aarhus C, Denmark}

\author{C. Klempt}
\affiliation{Institut f\"ur Quantenoptik, Leibniz Universit\"at Hannover, Welfengarten 1, 30167 Hannover, Germany}

\author{L. Santos}
\affiliation{Institut f\"ur Theoretische Physik, Leibniz Universit\"at Hannover, Appelstr. 2, 30167, Hannover, Germany}

\begin{abstract}
Spin-changing collisions may lead under proper conditions to the parametric amplification of matter waves 
in spinor Bose-Einstein condensates. Magnetic dipole-dipole interactions, although typically very weak in alkaline atoms, 
are shown to play a very relevant role in the amplification process. We show that these interactions may lead to a 
strong dependence of the amplification dynamics on the angle between the trap axis and the magnetic-field orientation. 
We analyze as well the important role played by magnetic-field gradients, which modify also strongly 
the amplification process. Magnetic-field gradients must be hence carefully controlled in future experiments, 
in order to observe clearly the effects of the dipolar interactions in the amplification dynamics.
\end{abstract}


\maketitle

\section{Introduction}

Spinor Bose-Einstein condensates (BECs), formed by atoms with various available Zeeman states, have attracted a large 
attention in recent years, mostly motivated by the rich physics resulting from the interplay between 
internal and external degrees of freedom. In addition to a wealth of possible 
ground-state phases~\cite{Ho1998,Ohmi1998,Koashi2000,Ciobanu2000}, the spinor 
dynamics has been at the focus of major interest~\cite{Chang2004,Schmaljohann2004}. 
This dynamics results from spin-changing collisions, which 
coherently re-distribute the populations among the different Zeeman states. 
Interestingly, spin-changing collisions are typically characterized by a very low energy scale much lower 
than the chemical potential in the condensate. As a result of that, the spinor dynamics in alkaline gases 
may be extraordinarily sensitive to other small energy scales.

Up to very recently, only short-range interactions have played a role in typical experiments in ultra-cold gases. 
Recent experiments have started to unveil the rich physics resulting from the dipole-dipole 
interactions (DDI)~\cite{baranov2008,lahaye2009}. 
This is particularly the case of Chromium, which presents a relatively large magnetic dipole moment, $\mu=6\mu_B$. 
Remarkable effects of the magnetic DDI have been reported in recent experiments on 
Chromium BEC~\cite{lahaye2007,koch2008,lahaye2008,metz2009}. Alkaline atoms, on the contrary, 
present a much lower magnetic dipole moment, $\mu=\mu_B/2$, and hence they are not usually expected to show 
any trace of the DDI unless short-range interactions are switched-off by means of Feshbach resonances~\cite{fattori2008,Pollack2009}.
However, as mentioned above, the spin-changing collisions in alkaline spinor BECs (in particular $F=1$ $^{87}$Rb) are remarkably 
low-energetic. As a result, spinor dynamics is very sensitive to magnetic DDI, in spite of the very low magnetic dipole 
moment. Recent experiments~\cite{vengalattore2008} have shown that the DDI may induce magnetization patterns 
in $F=1$ $^{87}$Rb BECs.

Spinor dynamics is particularly interesting for the case of condensates initially prepared in the $m=0$ Zeeman sublevel. 
In that case, spin-changing collisions may lead to correlated Einstein-Podolsky-Rosen (EPR) pairs 
in $m=\pm 1$~\cite{Duan2000,Pu2000}, in a 
process which closely resembles parametric down conversion in non-linear optics~\cite{Walls1994}. As a result, spinor condensates may 
act as parametric amplifiers of matter waves~\cite{Leslie2008,Klempt2009,Klempt2010}, 
opening interesting perspectives for the creation 
of non-classical states of matter based on spinor BECs. Recently we have shown that the interplay between trapping potential, 
quadratic Zeeman effect (QZE) and spin-changing collisions crucially determines the amplification gain~\cite{Klempt2009} and 
its sensitivity with respect to quantum spin fluctuations~\cite{Klempt2010}. 

In this paper we show that the amplification dynamics may be extremely sensitive to the DDI. 
As a result of that, the amplification of EPR-like pairs is strongly modified by the relative orientation between 
the applied magnetic field and the trap axis. We analyze in detail this dependence, as well as the effects 
of magnetic-field gradients. We show that these gradients must be carefully controlled, since uncontrolled gradients 
may obscure the expected DDI effects. We finally comment on experimental requirements.

The structure of the paper is as follows. In Sec.~\ref{sec:Hamiltonian} we present the system considered 
and the corresponding Hamiltonian. The linear regime is discussed in Sec.~\ref{sec:linear-regime}. An intuitive 
qualitative picture of the effects of the DDI in the amplification dynamics is discussed in Sec.~\ref{sec:Qualitative}. 
In Sec.~\ref{sec:Dynamics} we introduce the main formalism to analyze the amplification dynamics in the presence of DDI, 
whereas the corresponding numerical results are presented in Sec.~\ref{sec:Results}. The effects of the magnetic-field 
gradient are analyzed in Sec.~\ref{sec:Gradient}. Finally we discuss experimental requirements and summarize our conclusions in 
Sec.~\ref{sec:Conclusions}.

%%%%%%%%%%%%%%%

% HAMILTONIAN

\section{Hamiltonian}
\label{sec:Hamiltonian}

In the following we consider a spin-$1$ Bose gas (e.g. $F=1$ $^{87}$Rb), with Zeeman components $m = 0, \pm 1$,
confined in a dipole trap in the presence of an external magnetic field 
(which we assume as oriented along the $z$-axis). The system is described by the Hamiltonian $\hat H = \hat H_0 + \hat H_\text{sr} + \hat H_\text{dd}$. 
In this Hamiltonian, the single-particle physics is described by
\begin{equation}
  \hat H_0\! =\!\! \!\int\! d^3r\!\sum_m 
\hat \psi_m^\dagger (\vec r) \!\left [ -\frac{\hbar^2 \Delta}{2 M}\! +\! V (\vec r)\! +\! E_Z (m) \right ]\! \hat \psi_m (\vec r), 
\end{equation}
where $\hat \psi_m$ annihilates bosons with spin projection $m$. The trapping potential is of the form 
$V(\vec r) = \frac{M}{2} \Bigl( \omega_\perp^2 \bigl( x^2 + y'^2 \bigr) + \omega_\parallel^2 z'^2 \Bigr)$,  
where $M$ is the atomic mass, $\omega_\parallel \ll \omega_\perp$ are the trap frequencies (cigar-shaped trap), 
and $y' = \cos \vartheta \, y + \sin \vartheta \, z$, $z' = -\sin \vartheta \, y + \cos \vartheta \, z$, with 
$\vartheta$ the angle between the trap axis and the magnetic field orientation. This angle will play a crucial role in 
our discussion of the effects of the DDI. 

The Zeeman energy for the $m$ component is of the form:
\begin{equation}
  E_Z(m) \simeq \left( p + \nabla p \cdot \vec r \, \right)m + q m^2 ,
\end{equation}
where $p=g_L \mu_B B_0$ characterizes the linear Zeeman effect (LZE) for a homogeneous magnetic field $B_0$, 
with $g_L$ the Land\'e factor ($g_L = -1/2$ for $F=1$ $^{87}$Rb) and $\mu_B$ the Bohr magneton. The quadratic Zeeman 
effect (QZE) is characterized by a constant $q$, which in principle depends as $q = \mu_B^2 B_0^2 / (8 C_\text{hfs})$ on the 
hyperfine coupling strength $C_\text{hfs}$ ($ \approx h\times 3.4$ GHz for $^{87}$Rb), but may be also externally modified 
using optical or micro-wave dressing~\cite{Gerbier2006,Santos2007}. 
Additionally, we allow for a magnetic field gradient $B=B_0+{\vec \nabla} B\cdot {\vec r}$, 
leading to an energy shift $m {\vec \nabla} p \cdot \vec r$, which plays a relevant role below.

The short-range interactions are given by:
\begin{equation}
\hat H_{sr}\!=\! \frac{1}{2}\int d^3 r \!\! \sum_{m,m' \atop \bar m',\bar m} \!
\hat \psi_{m}^\dagger (\vec r) \hat \psi_{m'}^\dagger (\vec r) 
U_{m,m'}^{\bar m',\bar m}
\hat \psi_{\bar m'} (\vec r) \hat \psi_{\bar m} (\vec r),
\end{equation}
with $U_{m,m'}^{\bar m',\bar m}\equiv U_0\delta_{m,\bar m}\delta_{m',\bar m'} + 
U_1 \vec f_{{m} {\bar m}} \cdot \vec f_{{m'} {\bar m'}}$, 
where $\vec f_{{m} {m'}} = (f^x_{{m} {m'}}, f^y_{{m} {m'}}, f^z_{{m} {m'}})^T$, with $f_{x,y,z}$ the spin-$1$ matrices.  
$U_0 = (g_0 + 2 g_2) / 3$ and $U_1 = (g_2 - g_0) / 3$ are, respectively, the coupling constants 
for spin-independent and spin-dependent interactions, 
where $g_F = 4 \pi \hbar^2 a_F / M$, with $a_F$ the s-wave scattering length for the 
channel with total spin $F$. Note that the short-range interactions preserve the total spin projection, 
but this may be done in two crucially different ways, either by preserving the individual spin projections 
(spin-preserving collisions) or by modifying the individual projections while preserving the total one 
(spin-changing collisions).

Finally, the magnetic dipole-dipole interaction is given by:
\begin{eqnarray}
\hat H_{dd}&=& \frac{1}{2}\int d^3 r \int d^3r' \!\! \sum_{m,m' \atop \bar m',\bar m} \!
\hat \psi_{m}^\dagger (\vec r) \hat \psi_{m'}^\dagger (\vec r\,') \nonumber \\
&&W_{m,m'}^{\bar m',\bar m}(\vec r-\vec r\,')
\hat \psi_{\bar m'} (\vec r\,') \hat \psi_{\bar m} (\vec r),
\end{eqnarray}
with 
\begin{eqnarray}
W_{m,m'}^{\bar m',\bar m}(\vec r-\vec r\,')&\equiv& \frac{d^2}{|\vec r-\vec r\,'|^3}
\left[ \vec f_{{m} {\bar m}} \cdot \vec f_{{m'} {\bar m'}} \right\delimiter 0 \nonumber \\
&-& 3 \left\delimiter 0 \left( \vec f_{{m} {\bar m}} \cdot \vec u_r \right) 
\left( \vec f_{{m'} {\bar m'}} \cdot \vec u_r \right) \right], 
\end{eqnarray}
where 
$d^2 = \mu_0 g_L^2 \mu_B^2 / (4 \pi)$, $\mu_0$ is the vacuum permeability, and 
$\vec u_r = (\vec r - \vec r\,') / |\vec r - \vec r\,'|$. Contrary to the short-range 
interactions the DDI may violate, in principle, the conservation of the total spin projection 
(they may induce the equivalent of the Einstein-de Haas effect~\cite{Santos2006,Kawaguchi2006}). 
However, the associated change in LZE is typically, 
even for very low magnetic fields, orders of magnitude larger than any energy in the system and hence these 
spin-violating processes can be safely considered as suppressed.

Since short-range interactions preserve spin projection, and so do in practice DDI as well, the homogeneous LZE ($pm$) is preserved 
and it may be gauged out, playing no role in the dynamics discussed below. The same argument cannot be, however, employed with the 
magnetic-field gradient which may play a significant role in the spinor dynamics~\cite{Cherng2009}, as discussed in Sec.~\ref{sec:Gradient}.

%%%%%%%%%%%%%

% LINEAR REGIME

\section{Linear regime}
\label{sec:linear-regime}

In the following we are interested in the first stages (linear regime) 
of the spinor dynamics of a spin-$1$ BEC initially prepared in the 
$m=0$ sublevel, after quenching $q$ into the unstable regime.  
This dynamics, induced by spin-changing collisions, 
is characterized by the correlated creation of atomic pairs in $m=\pm 1$. 
In this section, and for the sake of simplicity, we do not consider 
magnetic-field gradients, which will be introduced in Sec.~\ref{sec:Gradient}. 

Before quenching $q$, the BEC is prepared in the $m=0$ component. The initial 
scalar wavefunction $\psi_0$ of the BEC in $m=0$ and the corresponding chemical potential $\mu$ 
may be obtained from the time-independent Gross-Pitaevskii equation:
\begin{equation} \label{scalar-GPE}
\left[ -\frac{\hbar^2}{2 M} \Delta + V (\vec r) + 
U_0 n_0 (\vec r) \right] \psi_0 (\vec r) = \mu \psi_0 (\vec r), 
\end{equation}
with $n_0 (\vec r) = |\psi_0(\vec r)|^2$, and $\int d^3 r |\psi_0(\vec r)|^2=N$.

The first stages of the spinor dynamics may be described by means of a Bogoliubov approximation:
\begin{equation} \label{field-operator}
(\hat \psi_1, \hat \psi_0, \hat \psi_{-1})^T = \left[ (0, \psi_0, 0)^T + (\delta \hat \psi_1, \delta \hat \psi_0, \delta \hat \psi_{-1})^T \right] e^{-i \mu t} .
\end{equation}
where we consider small fluctuations of the spinor field operator $\{\delta \hat \psi_m \}$, 
such that $|\psi_0|^2 \gg \sum_{m} \langle \delta \hat \psi_{m}^\dagger \delta \hat \psi_{m} \rangle$.

\subsection{Hamiltonian without dipole-dipole interactions}
\label{subsec:Hamiltonian-noDDI}

We consider first the Hamiltonian without DDI. Inserting~(\ref{field-operator}) into the grand canonical potential 
$\hat H_0+\hat H_{sr}-\mu \hat N$ (with $\hat N = \int d^3r \sum_m \hat \psi_m^\dag (\vec r) \hat \psi_m(\vec r)$ the total particle number), 
and keeping terms up to second order in $\delta \hat \psi_m$, we obtain an effective Hamiltonian for $\delta \hat \psi_{\pm 1}$
of the form:
\begin{eqnarray} \label{Hamiltonian-for-fluctuations}
\hat H_1 & = & \sum_{m = \pm 1} \int d^3 r \delta \hat \psi_{m}^\dagger \left( \hat H_{eff} + q \right ) 
\delta \hat \psi_m \nonumber \\
  & + & U_1 \int d^3 r\; n_0 \left( \delta \hat \psi_1^\dagger \delta \hat \psi_{-1}^\dagger + 
\delta \hat \psi_1 \delta \hat \psi_{-1} \right) .
\end{eqnarray}
Note that in the linear regime the fluctuations $\delta \hat \psi_{\pm 1}$ are decoupled from the density and phase 
fluctuations $\delta \hat \psi_0$ of the $m=0$ BEC (which may be excited during the preparation process).
In the previous expression we have introduced $\hat H_{eff} \equiv -\hbar^2 \Delta / 2 M + V_{eff} (\vec r)$, 
where
\begin{equation} \label{effective-trap}
  V_{eff} (\vec r) = V (\vec r) + (U_0 + U_1) n_0 (\vec r) - \mu
\end{equation}
may be understood as the effective trapping potential felt by the $\pm 1$-fluctuations on top of the $m=0$ BEC. It contains the mean-field potential $(U_0 + U_1) n_0$, which originates from spin-preserving collisions of $\pm 1$-atoms with the BEC in $m = 0$. 
Note that in the Thomas-Fermi regime $\mu=V(\vec r)+U_0n_0(\vec r)$. In that regime, $V_{eff}(\vec r)=U_1 n_0(\vec r)$ within 
the BEC region, and $V_{eff}=V(\vec r)-\mu$ outside.

The second line in Eq.~(\ref{Hamiltonian-for-fluctuations}) originates from spin-changing collisions, which 
convert $m = 0$ atoms into $\pm 1$-atom pairs and vice-versa. Interestingly, this process 
resembles parametric down conversion in optical parametric amplifiers~\cite{Walls1994}. Indeed, if the 
$m=0$ BEC is unstable after the quench of $q$, spin-changing collisions lead to an exponential amplification of the 
population in $m=\pm 1$~\cite{Leslie2008,Klempt2009,Klempt2010}.

\subsection{Dipole-dipole interactions}
\label{subsec:Hamiltonian-withDDI}

As mentioned above, spin-changing collisions are typically characterized by 
a very low energy scale. As a result, the spinor physics is highly sensitive to other very low energy scales, 
including the rather weak magnetic DDI in alkaline gases. This sensitivity has been recently demonstrated in experiments 
on the formation of spatial magnetization patterns in spinor Rb BECs~\cite{vengalattore2008}. As we show in following sections, 
also the exponential amplification of the population in $m=\pm 1$ following a quench in $q$ may be very significantly modified 
by the DDI.

Although there is no DDI contribution in the GP Eq.~(\ref{scalar-GPE}), since $\vec f_{00} = 0$,
there is however an important contribution to the effective Hamiltonian for $\delta \hat \psi_{\pm 1}$ 
which we obtain after inserting ~(\ref{field-operator}) in $\hat H_{dd}$ and linearizing:
\begin{eqnarray} \label{dipolar-Hamiltonian-for-fluctuations}
  \hat H_{1,dd} & = & \int d^3 r \int d^3 r' \psi_0 (\vec r) \psi_0 (\vec r\,') V_\text{dd} (\vec r - \vec r\,') \nonumber \\
  & & \times \biggl[ \delta \hat \psi_1^\dagger (\vec r) 
\delta \hat \psi_1 (\vec r\,') + \delta \hat \psi_{-1}^\dagger (\vec r) \delta \hat \psi_{-1} (\vec r\,') \nonumber \\
  & &  + \, \delta \hat \psi_1^\dagger (\vec r) \delta \hat \psi_{-1}^\dagger (\vec r\,') + \delta \hat \psi_1 (\vec r) \delta \hat \psi_{-1} (\vec r\,') \biggr] 
\end{eqnarray}
with $V_\text{dd} (\vec r) = \frac{d^2}{2|\vec r|^5} \left( 3 z^2-|\vec r|^2\right)$.
In Eq.~(\ref{dipolar-Hamiltonian-for-fluctuations}) we have neglected terms related to scattering processes which do not preserve the 
total spin projection since, as mentioned in Sec.~\ref{sec:Hamiltonian}, the associated change in LZE suppresses 
spin-violating processes even for very low magnetic fields. Note that the third line of $\hat H_{1,dd}$ 
contains as well a parametric amplification term.

%%%%%%%%%%%%%%%%%%%%%%%%%%%%%

% INTUITIVE PICTURE

\section{Qualitative picture of the effect of the dipole-dipole interactions on the amplification dynamics}
\label{sec:Qualitative}

The effects of the DDI on the amplification dynamics may be qualitatively understood from a simplified 
model. In homogeneous space ($V=0$, constant $n_0$, $\mu=U_0n_0$) we may introduce the Fourier transform
$\hat \eta_m (\vec k) = \int d^3r \, \delta \hat \psi_m(\vec r) e^{-i \vec k \cdot \vec r}$, which allows to write $\hat H_1$ 
in the simplified form~\cite{Lamacraft2007}: $H_1^{hom}=\int  \hat h_k d^3 k/(2\pi)^3$, where
\begin{eqnarray}
\hat h_k&=& 
\sum_{m = \pm 1} \left (E_k+q-q_{cr}\right )\eta_{m}^\dag(\vec k)\eta_{m}(\vec k)\nonumber \\
&-& q_{cr}\left ( 
\hat\eta_{1}^\dag(\vec k)\hat\eta_{-1}^\dag(-\vec k)+\hat\eta_{1}(\vec k)\hat\eta_{-1}(-\vec k)\right )
,
\end{eqnarray}
with $E_k=\hbar^2k^2/2M$ and $q_{cr}=-U_1 n_0$. 
Note that for $F=1$ $^{87}$Rb $U_1<0$, and hence $q_{cr}>0$ (we consider in the following this case, 
although for the $F=2$ case the sign is the opposite).
This Hamiltonian, which may be easily diagonalized for each momentum $\vec k$, possesses 
eigenenergies of the form 
\begin{equation}
\lambda^\pm(\vec k)=\sqrt{(E_k+q-q_{cr})^2-q_{cr}^2}. 
\end{equation}
Note that if $\text{Im}(\lambda^\pm (\vec k)) > 0$ for some eigenenergy, then 
there is an exponential growth of spin excitations, which leads to a correlated pair creation in $m=\pm 1$. 
This instability is best characterized by the instability rate $\Lambda={\rm max} \{{\rm Im}(\lambda_\pm(\vec k)) \}$.
It is straightforward to show that the $m=0$ BEC becomes unstable for $q < 2 q_{cr}$. The instability rate rises 
between $q_{cr} < q < 2 q_{cr}$ acquiring its maximal value $\Lambda = q_{cr}$ at $q = q_{cr}$. For $q < q_{cr}$, $\Lambda = q_{cr}$ in the homogeneous case 
(in the inhomogeneously trapped case the instability rate $\Lambda$ presents significant modulations which are responsible 
of the multi-resonant $q$-dependence of the amplification dynamics recently observed experimentally~\cite{Klempt2009}). 

In the following we apply a similar formalism to the DDI term $\hat H_{1,dd}$. 
We introduce the Fourier transformation of $V_{dd}(\vec r)$ to obtain $\tilde V_{dd}(\vec k)=U_{dd} (1-3\cos^2\theta)$, where 
$U_{dd}=2\pi d^2/3$ and $\theta$ is the angle between $\vec k$ and the dipole orientation ($z$ axis). 
Using convolution theorem, and since we assume $n_0$ as constant, we may then rewrite $\hat H_{1,dd}$ in the form:
\begin{eqnarray}
\hat H_{1,dd}^{hom}&=& \int \frac{d^3k}{(2\pi)^3} n_0 \tilde V_{dd}(\vec k) 
\left \{ \sum_{m=\pm 1} \hat\eta_m^\dagger (\vec k) \hat\eta_m(\vec k) \right\delimiter 0 \nonumber \\
&+&\left\delimiter 0 \hat\eta_{1}^\dagger (\vec k) \hat\eta_{-1}(-\vec k)^\dag+\hat\eta_{1}(\vec k)\hat\eta_{-1}(-\vec k)
\right\}.
\end{eqnarray}
Note that adding $H_{1,dd}$ to $H_1$ results in a similar form as that of $\hat H_1$ but with an effective 
$q_{cr}^{eff}(\theta)=q_{cr}-n_0 \tilde V_{dd}(\vec k)$. Note that, due to the 
anisotropy of the DDI, $q_{cr}^{eff}$ depends on the angle $\theta$. 
The effects of the trap geometry may be qualitatively understood from this $\theta$ dependence.
For an axisymmetric trap the dominant momenta are those along the 
tightest direction. If the dipole orientation is perpendicular to the trap axis, then 
the dominant $\vec k$ will then be those with $\theta=0$, and hence $q_{cr}^{eff} \simeq q_{cr} + 2 n_0 U_{dd}$. On the contrary, if the dipole orientation 
is parallel to the trap axis, the dominant momenta will be those with $\theta=\pi/2$, and 
$q_{cr}^{eff}\simeq q_{cr} - n_0 U_{dd}$. Since $\Lambda \simeq q_{cr}^{eff}$, we hence expect 
an enhancement of the instability for a magnetic field orientation perpendicular to the trap axis, and 
a reduced instability for a parallel orientation. Although the DDI in alkaline atoms is typically very weak, 
the spin-changing collisions are very weak as well. In particular, in $F=1$ $^{87}$Rb the strength of the DDI is quite significant 
compared to the strength of the spin-changing collisions, $|U_{dd}/U_1| \approx 0.2 $. As a result, 
the DDI modification of the instability rate is expected to lead to a marked orientation dependence 
of the amplification dynamics. In the following sections we show that this is indeed the case 
when considering realistic trapped cases.

%%%%%%%%%%

% FIGURE 1
\begin{figure}%[t]
\includegraphics[width = 7.5cm]{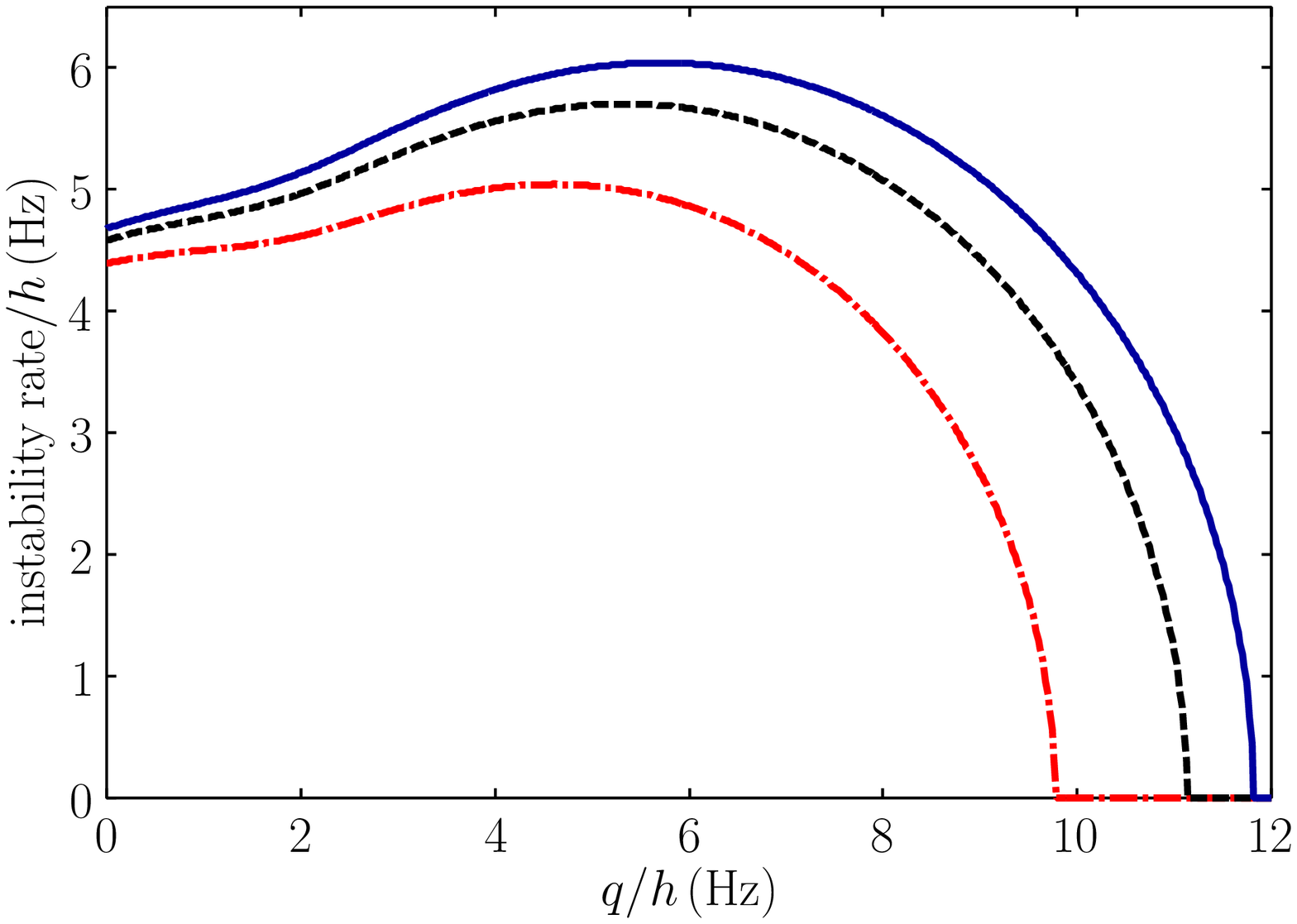}
\caption{(Color online) Instability rate $\Lambda$ as a function of $q$ for $N=10^5$, $\omega_\perp = 2 \pi \times 200\,$Hz and $\omega_\parallel = 2 \pi \times 40\,$Hz, for the case of no DDI (black, dashed), $\theta=\pi/2$ (blue, solid) and $\theta=0$ (red, dot-dashed).}  \label{fig:1}
\end{figure}

%%%%%%%%%%%%%

% TRAPPED CASE

\section{Amplification dynamics in trapped dipolar condensates}
\label{sec:Dynamics}

Although the homogeneous model discussed before allows for a simplified intuitive understanding of the major effects 
of the DDI in the amplification process, a quantitative analysis of realistic experimental situations may be 
just achieved by properly considering the inhomogeneous trapping, and the corresponding inhomogeneous density $n_0(\vec r)$ 
of the $m=0$ BEC. In this section we introduce the basic formalism which we follow for the analysis of the 
amplification dynamics characterizing the spinor physics in the linear regime.

The analysis of the spinor dynamics is significantly simplified by considering the eigenfunctions and eigenenergies of 
$\hat H_{eff}$, $\hat H_{eff}\phi_n(\vec r)=\epsilon_n\phi_n(\vec r)$, and expanding the field operators in the 
basis of these eigenstates $\delta \hat \psi_m(\vec r) = \sum_n \phi_n(\vec r) \hat a_{n m}$. We may then rewrite:
\begin{eqnarray} \label{Hamiltonian-for-fluctuations-b}
\hat H_1+H_{1,dd}& = & \nonumber \\
&&\!\!\!\!\!\!\!\!\!\!\!\!\!\!\!\!\!\!\!\!\!\!\!\!\!\!\!\!\!\!\!\!\!\!\!\!\!\sum_{n n'} ((\epsilon_n + q)\delta_{n n'}+B_{n n'}) \sum_{m=\pm 1}\hat a_{n m}^\dagger \hat a_{n' m} \nonumber \\
  &&\!\!\!\!\!\!\!\!\!\!\!\!\!\!\!\!\!\!\!\!\!\!\!\!\!\!\!\!\!\!\!\!\!\!\!\!\! + \sum_{n n'} (A_{n n'}+B_{n n'}) \left( \hat a_{n 1}^\dagger \hat a_{n' -1}^\dagger + \hat a_{n 1} \hat a_{n' -1} \right),
\end{eqnarray}
where $A_{n n'} = U_1 \int d^3 r \, n_0 \phi_n \phi_{n'}$ characterizes the effects of the short-range spin-changing collisions, whereas the effects of the DDI are given by
\begin{equation} \label{DDI-integrals}
  B_{n n'}\!=\!\! \int\! d^3 r \int d^3 r' F_n(\vec r)V_{dd} (\vec r - \vec r\,')F_{n'}(\vec r'),
\end{equation}
where $F_n(\vec r)=\psi_0 (\vec r)\phi_n(\vec r)$. The matrix elements $B_{n n'}$ are 
most efficiently calculated in $\vec k$-space according to
\begin{equation}
B_{n n'} = \int \frac{d^3 k}{(2\pi)^3} {\tilde F}_n (\vec k) \tilde V_{dd}(\vec k) {\tilde F}_{n'} (\vec k)
\end{equation}
where $\tilde F(\vec k)$ is the Fourier transform of $F(\vec r)$. 

Eq.~(\ref{Hamiltonian-for-fluctuations-b}) is solved by the multimode Bogoliubov ansatz
\begin{equation} \label{quasiparticles}
  \hat \alpha_\nu^\pm = \sum_n \left( u_{\nu n}^\pm \hat a_{n 1} + v_{\nu n}^\pm \hat a_{n -1}^\dagger \right), 
\end{equation}
where $\hat\alpha_{\nu}^\pm$ satisfy
$ \left[ \hat \alpha_\nu^\pm, \hat H_1 + H_{1,dd} \right] = \lambda_\nu^\pm \hat \alpha_\nu^\pm$, which leads to the eigenvalue equation:
\begin{equation} \label{eigenvalue-equation}
  {\mathbf C}\cdot  \begin{pmatrix} \vec u_\nu^\pm  \\ \vec u_\nu^\pm  \end{pmatrix} = 
\lambda_\nu^\pm \begin{pmatrix}\vec u_\nu^\pm  \\ \vec u_\nu^\pm \end{pmatrix},
\end{equation}
where $\vec u_\nu^{\pm\; T}\equiv\{u_{\nu 1}^\pm,u_{\nu 2}^\pm,\dots \}$ (and similarly for $\vec v_\nu^\pm$)  
and  
\begin{equation} \label{matrix-C}
    \!{\mathbf C}=\begin{bmatrix}
    {\mathbf E} + q {\mathbf 1} +{\mathbf B}& -{\mathbf A}+{\mathbf B}) \\
    {\mathbf A}+{\mathbf B} & -{\mathbf E} - q {\mathbf 1} -{\mathbf B}
  \end{bmatrix},
\end{equation}
with $E_{nn'} = \epsilon_n \delta_{nn'}$, ${\mathbf 1}$ the identity matrix, and ${\mathbf A}$ ($\mathbf B$) the matrix with components $A_{n n'}$ ($B_{n n'}$). 
From the Heisenberg equations of motion
$\hat \alpha_\nu^\pm (t) = \hat \alpha_\nu^\pm (0) e^{-i \lambda_\nu^\pm t / \hbar}$.
Note that, as for the homogeneous case, if $\text{Im}(\lambda_\nu^\pm) > 0$ for some eigenenergy, then 
there is an exponential growth of correlated pairs in $m=\pm 1$. 
As for the homogeneous case, this instability is best characterized by the instability rate $\Lambda={\rm max} \{{\rm Im}(\lambda_\nu^\pm) \}$.

The time evolution of $\hat a_{n,\pm 1}$ is then easily obtained 
\begin{equation} \label{trap-state-evolution}
  \begin{bmatrix} \{\hat a_{n 1} (t)\} \\ \{\hat a_{n -1}^\dagger (t)\} \end{bmatrix} = {\mathbf U}(t) \begin{bmatrix} \{ \hat a_{n 1} (0)\} \\ \{ \hat a_{n -1}^\dagger (0)\} \end{bmatrix} ,
\end{equation}
with ${\mathbf U}={\mathbf M}^{-1} e^{-i {\mathbf \Lambda} t / \hbar} {\mathbf M}$, 
where ${\mathbf M}$ is the matrix of eigenvectors obtained after solving Eq.~(\ref{eigenvalue-equation}) and ${\mathbf \Lambda}$ the corresponding diagonal matrix of 
eigenvalues.

As mentioned above, the atoms are initially prepared in the $m=0$ sublevel. However, a slightly imperfect preparation may lead 
to a non-zero population of $N_s$ $m=\pm 1$ pairs in the original BEC~\cite{Klempt2010}. 
These spurious atoms (which from now are called {\em classical seed})
share the same wavefunction $\psi_0(\vec r)$ as the $m=0$ atoms. Denoting $\chi_n=\int d^3 r \psi_0(\vec r) \phi_n (\vec r) / \sqrt{N}$, 
we may then easily express the population $P_m=\sum_n \langle \hat a_{n,m}^\dag \hat a_{n,m}\rangle $ in the form
$P_m(t)=P_{C}(t)+P_Q(t)$, where 
\begin{equation}
P_{C} (t) = N_s \vec \chi \cdot \left( {\mathbf O}^\dagger {\mathbf O} + \widetilde {\mathbf O}^\dagger \widetilde {\mathbf O} \right) \vec \chi,
\label{eq:PC}
\end{equation}
denotes the population triggered by the classical seed, and 
\begin{equation}
P_{Q} (t) = \text{Tr} \left( \widetilde {\mathbf O}^\dagger \widetilde {\mathbf O} \right),
\label{eq:PQ}
\end{equation}
denotes the population induced by quantum fluctuations (i.e. when $N_s=0$). In the previous expressions, 
the matrices ${\mathbf O}$ and $\widetilde {\mathbf O}$ are the upper left and upper right part of the time 
evolution matrix ${\mathbf U}(t)$ and $\vec \chi = (\chi_1, \chi_2, ...)^T$.

%%%%%%%%%%%%%

% NUMERICAL RESULTS

\section{Dipole-induced orientation-dependence of the amplification dynamics}
\label{sec:Results}

In this section we employ the formalism discussed in Sec.~\ref{sec:Dynamics} to study the effects of the DDI 
in the amplification dynamics. We shall show that due to the DDI the amplification may be markedly dependent on the 
relative orientation between the trap axis and the external magnetic field.

In our numerical calculations we have considered realistic experimental conditions, with $N=10^5$ $F=1$ $^{87}$Rb atoms 
in a cigar-shaped harmonic potential with $\omega_\perp=2\pi\times 200$~Hz, and $\omega_\parallel=2\pi\times 40$~Hz.
As mentioned above, we consider the atoms as initially prepared in $m=0$ with possibly an initial spurious classical 
seed (which we typically consider as $N_s=2$, a typical value expected from previous experimental results~\cite{Klempt2010}). 
At $t=0$ the QZE energy $q$ is set to a given value within the instability regime. We monitor the subsequent evolution of the 
populations $P_{\pm 1} (t)$ obtained from Eqs.~(\ref{eq:PC}) and (\ref{eq:PQ}) as a function of $q$ and the relative 
angle $\theta$ between the trap axis and the external magnetic field.

Fig.~\ref{fig:1} shows the dependence of the instability rate $\Lambda$ as a function of $q$ for $\theta=0$, $\theta=\pi/2$ and 
without DDI. Note that for all cases, the instability rate experiences a maximum contrary to the 
homogeneous case. This maximum is induced by the inhomogeneous harmonic trapping and leads to marked resonances in the 
$q$-dependence of the amplification dynamics, as discussed in Ref.~\cite{Klempt2009}. However, $\Lambda$ clearly 
depends on the trap orientation confirming indeed the intuitive qualitative picture discussed 
in Sec.~\ref{sec:Qualitative}. When trap axis and magnetic field are aligned $\Lambda$ decreases compared to the non-dipolar case, 
whereas the opposite is true when the magnetic field is oriented perpendicular to the trap axis. Note as well that, also as 
expected from the qualitative picture of Sec.~\ref{sec:Qualitative}, the instability region is shifted towards 
lower $q$ values in the parallel configuration, and towards larger $q$ values in the perpendicular one.

%%%%%%%%%%

% FIGURE 2
\begin{figure}%[t]
\includegraphics[width = 7.5cm]{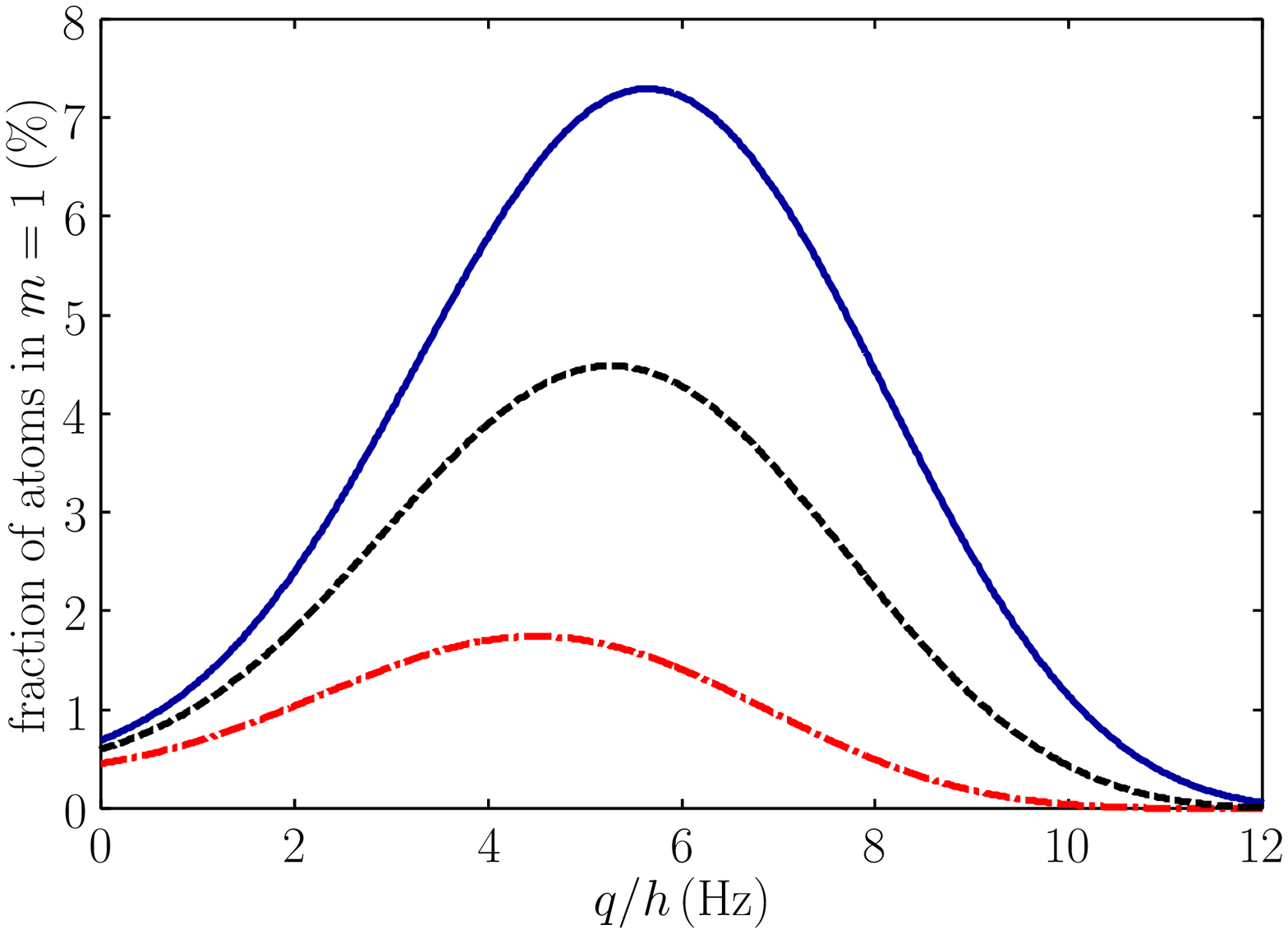}
\caption{(Color online) Fraction of atoms transfered into $| \pm 1 \rangle$ after 115\,ms as a function of $q$ for $N_s=2$ and the 
same parameters as Fig.~\ref{fig:1}, and for the case of no DDI (black, dashed), $\theta=\pi/2$ (blue, solid) and $\theta=0$ (red, dot-dashed).}
\label{fig:2}
\end{figure}

%%%%%%%%%%

% FIGURE 3

\begin{figure}%[t]
\includegraphics[width = 7.5cm]{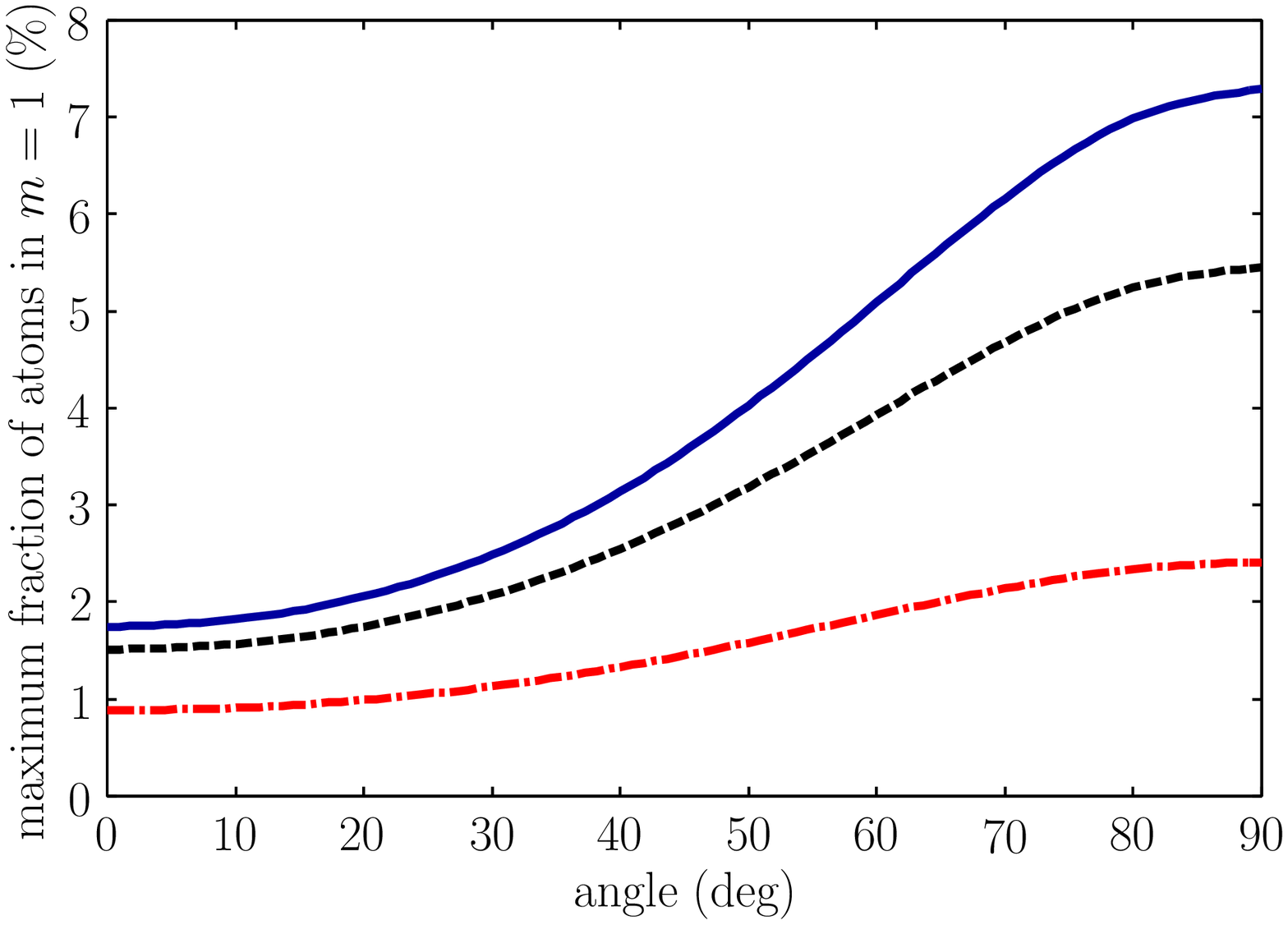}
\caption{(Color online) Maximal population transfer after 115\,ms as a function of $\theta$ (same parameters as Fig.~\ref{fig:2}) 
for different magnetic field gradients $\nabla B=0$ (blue, solid), $5\,$mG$/$cm (black, dashed) and $10\,$mG$/$cm (red, dot-dashed).}
\label{fig:3}
\end{figure}

%%%%%%%%%%%%%

This modified instability rate translates into a significantly distorted pair-creation dynamics, due to the 
exponential nature of the parametric amplification. Fig.~\ref{fig:2} shows the transferred fraction 
$P_{\pm 1}(t)/N$ after $t=115$~ms as a function of $q$ for different values of $\theta$. As expected 
from the form of $\Lambda$ we observe the appearance of a maximum for all $\theta$, which 
is slightly shifted (by approximately $1$~Hz) towards lower $q$ when $\theta$ is shifted from $\pi/2$ to $0$. 
However, this maximum is approximately four times as large for $\theta=\pi/2$ 
than for $\theta=0$. The dependence of the amplification on $\theta$ is very clearly observable in the $\theta$-dependence of the maximum of $P_{\pm 1}$ (again at $t=115$~ms) shown in Fig.~\ref{fig:3}. Note that the maximum grows monotonically from $\theta=0$ to $\theta=\pi/2$.

%%%%%%%%%%%%%

 % MAGNETIC FIELD GRADIENTS

\section{Effects of magnetic field gradients on the amplification dynamics}
\label{sec:Gradient}
As mentioned in Sec.~\ref{sec:Hamiltonian} the homogeneous LZE plays typically no role in the spinor dynamics 
(only at very low magnetic fields $B<1$~mG the DDI could induce the equivalent of the 
Einstein-de Haas effect~\cite{Santos2006,Kawaguchi2006}, 
and in this case the residual LZE could play a role). However, magnetic field gradients cannot be gauged out, 
and may play a relevant role in the spinor physics~\cite{Cherng2009}. In this section, we analyze 
the effects that these gradients may have on the amplification dynamics. We shall show that even 
relatively weak gradients may have a significant effect on the amplification process.

Although magnetic-field gradients do not affect the GP equation for the $m=0$ BEC, there is indeed a contribution to the 
effective Hamiltonian for $\delta \hat \psi_{\pm 1}$ in the linear regime:
\begin{equation}
  \hat H_\text{1,gr} = \nabla p \cdot \int d^3 r \left( \delta \hat \psi_1^\dagger \vec r \, \delta \hat \psi_1 - 
\delta \hat \psi_{-1}^\dagger \vec r \, \delta \hat \psi_{-1} \right), 
\end{equation}
which may be straightforwardly implemented into the matrix $\mathbf C$ of the eigenvalue equation~(\ref{eigenvalue-equation}),
\begin{equation} \label{matrix-C}
    {\mathbf C}=\begin{bmatrix}
    {\mathbf E} + q {\mathbf 1} +{\mathbf B}+{\mathbf D}& -{\mathbf A}+{\mathbf B}) \\
    {\mathbf A}+{\mathbf B} & - {\mathbf E} - q {\mathbf 1} - {\mathbf B}+{\mathbf D}
  \end{bmatrix},
\end{equation}
with $D_{n n'} = \nabla p \cdot \int d^3 r \, \phi_n \vec r \, \phi_{n'}$.

The magnetic-field gradients have two main effects. On one side, they modify the effective potential $V_{eff}(\vec r)$ 
in a different way for $m=1$ than for $m=-1$. This reduces the overlap of the $m=\pm 1$ atom clouds with the $m=0$ BEC 
and hence the scattering mediated transfer. On the other side, atoms placed at different locations experience 
different Larmor precession frequencies. Although this does not affect the local short-range interactions, it does 
modify the non-local DDI. For large-enough gradients this may lead to a time-averaged DDI~\cite{Cherng2009}.  
For weak gradients, as those considered below, the explicit time dependence induced by the gradients must be considered.

%%%%%%%%%%

% FIGURE 4

\begin{figure}%[t]
\includegraphics[width = 7.5cm]{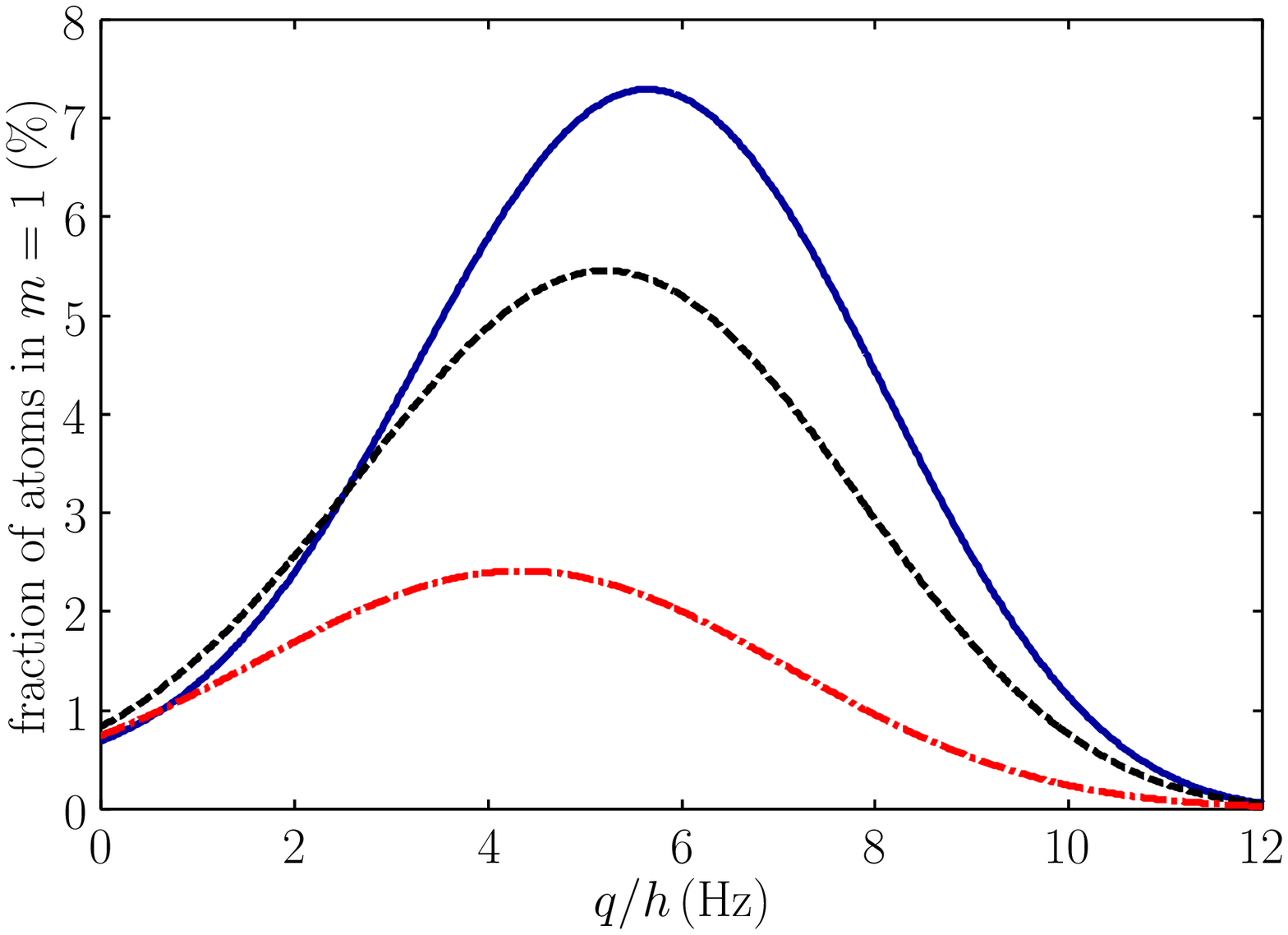}
\caption{(Color online) Fraction of atoms in $\pm 1$ after a time 115\,ms 
(same parameters as Fig.~\ref{fig:2}) as a function of $q$ 
for $\theta=\pi/2$, and for gradients $\nabla B=0$ (blue, solid), $5\,$mG$/$cm (black, dashed) and $10\,$mG$/$cm (red, dot-dashed).}
\label{fig:4}
\end{figure}

%%%%%%%%%%%%%

Parametric amplification is handicapped by the presence of gradients as a result of these two combined effects. 
Fig.~\ref{fig:4} shows the combined effect of the DDI and the magnetic-field gradient 
along the weak trap axis for $\theta=\pi/2$. As expected, we obtain a reduction of the transfer maximum with increasing gradient 
and a shift of its position to lower $q/h$ by approximately $1.5$~Hz for a gradient of $10$~mG/cm.
Hence the transfer maximum is shifted down and to lower $q$ with decreasing $\theta$ 
and increasing gradient. As shown in Fig.~\ref{fig:3}, in the presence of a magnetic field gradient the maximum 
of $P_{\pm 1}(t)$ shows also a marked $\theta$ dependence.

Hence, even rather weak gradients ($<10$~mG/cm) may strongly modify the amplification dynamics, an effect which is 
enhanced by the presence of the DDI. Although as mentioned above, the $\theta$-dependence should reveal also in the presence 
of gradients the effects of the DDI, slight variations of the magnetic field gradients (of the order of few mG/cm)  
when changing the magnetic-field orientation with respect to the trap axis must be very carefully controlled. 
This is indeed a crucial point, since  otherwise, accidental $\theta$-dependences of the magnetic-field 
gradients may obscure the physical $\theta$-dependence characteristic of the DDI.

\section{Discussion}
\label{sec:Conclusions}

\subsection{Experimental requirements}
\label{subsec:experiment}

In the following we outline the general requirements to perform an experiment with $F=1$ $^{87}$Rb to probe the 
theoretical results discussed above. A $m=0$ BEC must be prepared in a crossed dipole trap following the same procedure 
detailed in Refs.~\cite{Klempt2009,Klempt2010}. The state preparation requires particular attention, 
since remaining atoms in $m\neq0$ may strongly alter the experimental result. 
Previous experiments have shown that the number of atoms in $m\neq0$ states can be suppressed to $N_s\approx2$ 
by briefly applying a strong magnetic field gradient to purify the system~\cite{Klempt2010}. 
Due to the nature of the QZE in $F=1$ $^{87}$Rb, magnetic fields can be used to access positive values of $q$, as required above.

As discussed in Sec.~\ref{sec:Gradient}, the most significant requirement compared to previous experiments is related 
to the suppression of magnetic field gradients which could obscure the dipolar effects. Appropriate experiments should be 
carefully designed to minimize all sources of field gradients from the vicinity of the atomic sample 
(alternatively a magnetic shield could be placed around the sample). In state of the art precision measurements, 
field gradients are commonly suppressed below 1mG/cm~\cite{Gauguet2009}, which is sufficient to realize the mandatory experimental conditions (see Sec.~\ref{sec:Gradient}).

Fig.~\ref{fig:2} shows that the resonant spin transfer to the $m=\pm 1$ states depends strongly on the relative orientation 
of the weak trap axis and the external magnetic field. Since it is difficult to change the orientation 
of a dipole trap while maintaining its trapping potential, experiments must be designed to vary the 
orientation of the external magnetic field. In this sense, two sets of Helmholtz coils are necessary to provide 
a homogeneous external magnetic field. One of them should be placed along the weak axis of the 
trapping potential to realize the $\theta=0$ configuration and another one along one of the strong axis 
to realize the $\theta=\pi/2$ case. Both magnetic fields have to be calibrated, preferentially using 
precision microwave spectroscopy between the ground state hyperfine manifolds of $^{87}$Rb.  
Such an experimental apparatus would also allow for a rotation of the field, since 
the currents in the two Helmholtz coils could be adjusted to obtain a relative angle $\theta$. In this way, it should be 
possible to perform a measurement analogous to that discussed in Fig.~\ref{fig:3}. Finally, additional magnetic field 
gradients can be applied along both magnetic field directions to observe the suppression shown in Fig.~\ref{fig:4}. 

\subsection{Summary}
\label{subsec:summary}

We have shown that, in spite of the very small magnetic moment, the magnetic DDI may lead to a 
strong modification of the amplification dynamics in $F = 1$ $^{87}$Rb due to the low-energy scale of the 
spin-changing collisions. We have shown that the DDI induce a very marked dependence of the amplification gain with 
respect to the relative orientation between magnetic-field direction and trap axis. If both directions are perpendicular 
to each other the amplification dynamics is much faster than for the parallel configuration. Remarkably, the number 
of transferred atoms into $m=\pm 1$ may increase for $F = 1$ $^{87}$Rb for a fixed holding time of around $100$ ms by a factor over $400\%$ when turning from a 
parallel to a perpendicular configuration. We have shown as well that magnetic field gradients may also significantly modify 
the amplification dynamics, both due to their effects on the trapping and on the DDI. As a result, magnetic-field 
gradients must be carefully controlled, since uncontrolled changes in the gradient when turning the 
magnetic field orientation may obscure the orientation dependence of the DDI effects on the amplification.
This demands specific requirements for future experiments.

\acknowledgments
We acknowledge support from the Centre for Quantum Engineering and Space-Time Research QUEST, from the Deutsche Forschungsgemeinschaft (SFB 407), and the European Science Foundation (EuroQUASAR).

\bibliographystyle{prsty}

\end{document}